\documentclass[times,twocolumn,final]{elsarticle}

\usepackage{framed,multirow}

\usepackage{latexsym}
\usepackage{graphicx}
\usepackage{amsmath}
\usepackage{amssymb}
\usepackage{subcaption}
\usepackage{booktabs}
\usepackage{multirow}
\usepackage{colortbl}
\usepackage{float}
\usepackage{siunitx}

\usepackage{url}
\usepackage{xcolor}

\usepackage{hyperref}

\definecolor{newcolor}{rgb}{.8,.349,.1}

\begin{document}

\begin{frontmatter}

\title{Deep-LIBRA: Artificial intelligence method for robust quantification of breast density with independent validation in breast cancer risk assessment}%

\author[1]{Omid Haji Maghsoudi}
\ead{o.maghsoudi@gmail.com}
\author[1]{Aimilia Gastounioti}
\author[2]{Christopher Scott}
\author[1]{Lauren Pantalone}
\author[2]{Fang-Fang Wu}
\author[1]{Eric A. Cohen}
\author[2]{Stacey Winham}
\author[1]{Emily F. Conant}
\author[2]{Celine Vachon}
\author[1]{Despina Kontos}
\ead{despina.kontos@pennmedicine.upenn.edu}

\address[1]{Department of Radiology, University of Pennsylvania, Philadelphia, 19104, PA, USA}
\address[2]{Department of Health Sciences Research, Mayo Clinic, Rochester, 55905, MN, USA}

\begin{abstract}
Breast density is an important risk factor for breast cancer that also affects the specificity and sensitivity of screening mammography. Current federal legislation mandates reporting of breast density for all women undergoing breast screening. Clinically, breast density is assessed visually using the American College of Radiology Breast Imaging Reporting And Data System (BI-RADS) scale. Here, we introduce an artificial intelligence (AI) method to estimate breast percentage density (PD) from digital mammograms. Our method leverages deep learning (DL) using two convolutional neural network architectures to accurately segment the breast area. A machine-learning algorithm combining superpixel generation, texture feature analysis, and support vector machine is then applied to differentiate dense from non-dense tissue regions, from which PD is estimated. Our method has been trained and validated on a multi-ethnic, multi-institutional dataset of 15,661 images (4,437 women), and then tested on an independent dataset of 6,368 digital mammograms (1,702 women; cases=414) for both PD estimation and discrimination of breast cancer. On the independent dataset, PD estimates from Deep-LIBRA and an expert reader were strongly correlated (Spearman correlation coefficient = 0.90). Moreover, Deep-LIBRA yielded a higher breast cancer discrimination performance (area under the ROC curve, AUC = 0.611 [95\% confidence interval (CI): 0.583, 0.639]) compared to four other widely-used research and commercial PD assessment methods (AUCs = 0.528 to 0.588). Our results suggest a strong agreement of PD estimates between Deep-LIBRA and gold-standard assessment by an expert reader, as well as improved performance in breast cancer risk assessment over state-of-the-art open-source and commercial methods.
\end{abstract}

\end{frontmatter}


\section{Introduction}
Studies have shown that the relative amount of breast fibroglandular tissue, breast percent density (PD), not only limits the sensitivity of screening mammography but is also an independent breast cancer risk factor \cite{brentnall2018long, engmann2017population, freer2015mammographic}. Breast density can be estimated from full-field digital mammography (FFDM) and is most commonly assessed in the clinic by visual grading into one of the four categories defined by the  American College of Radiology BI-RADS \cite{d2003breast}. However, BI-RADS density assessment is highly subjective and does not provide a quantitative, continuous measure of PD, which would allow for more refined risk stratification and assessment of changes \cite{irshad2016effects, sprague2016variation}.

Automated quantitative PD measurement from FFDM is available through commercially available software  \cite{iCAD, hartman2008volumetric, regini2014radiological} and research-based tools \cite{anitha2017dual, czaplicka2011automatic, ferrari2004automatic, keller2012estimation, kwok2004automatic, li2013pectoral, mustra2013robust, mustra2016review, nagi2010automated, rampun2017fully, shi2018hierarchical, taghanaki2017geometry}. Although these tools are useful, important limitations persist. Several commercially available packages \cite{hartman2008volumetric, regini2014radiological} calculate PD based on x-ray beam interaction models. These packages make assumptions based on specific metadata to simplify various estimates, including identifying the fatty tissue. Therefore, these assumptions can lead to inaccurate estimates, especially when the required metadata is lacking. Moreover, commercial tools are costly, making them inaccessible for general use, and can require a specific image format or a particular FFDM device manufacturer. On the other hand, with a few exceptions, such as LIBRA  \cite{keller2012estimation, gastounioti2020evaluation}, research-based methods are not freely available, making it challenging to adopt such tools broadly and rigorously compare their performances. Research-based tools have also been developed using small, single-institution datasets, and lack independent validation \cite{anitha2017dual, keller2012estimation, li2013pectoral,  shi2018hierarchical}.

In general, the key computational steps for automated PD quantification from FFDM are image background removal; identification of the pectoralis muscle; and segmentation of the dense tissue areas. Background removal consists of identifying the air and extraneous objects (paddles, markers, rings, etc.) to remove these areas from density calculations. Similarly, the pectoralis muscle must be removed from the area to be processed, which can be challenging due to anatomic variation of the pectoralis muscle and the extension of dense glandular tissue which often superimposes over the pectoralis muscle in the axillary tail. To simplify its delineation, the pectoralis muscle has typically been modeled as a straight line \cite{ferrari2004automatic, keller2012estimation, kwok2004automatic, mustra2016review} or a curve \cite{mustra2013robust}), and such simplified assumptions can lead to inaccurate PD estimation. Most crucial to PD calculation is the segmentation of dense versus non-dense tissue. Most methods for this to date \cite{anitha2017dual, keller2012estimation, zhou2001computerized} are relatively simplistic, leading to over- or underestimating the amount of dense tissue. Like LIBRA, but unlike other techniques, Deep-LIBRA employs textural features in dense-tissue segmentation, but, unlike the earlier tool, incorporates this information into the artificial intelligence (AI) approach.

AI, including deep learning, has shown great potential in substantially improving medical image segmentation, classification, risk assessment and breast cancer detection \cite{becker2017deep, hamidinekoo2018deep, kaul2019focusnet, kontos2019can, kooi2017large, lehman2018mammographic, mohamed2018deep, mortazi2018automatically, murugesan2019psi, rodriguez2018detection, ronneberger2015u, wang2016discrimination, yala2019deep}. The U-Net deep learning architecture is the most widely used convolutional neural network (CNN) architecture for segmentation tasks; it has been used in breast cancer imaging \cite{dalmics2017using, hamidinekoo2018deep, kallenberg2016unsupervised, ma2019automated, rodriguez2018pectoral} and segmenting the pectoralis muscle in both FFDM \cite{ma2019automated} and digital breast tomosynthesis \cite{rodriguez2018pectoral}. A unique advantage of the U-Net is that it offers the superior accuracy of deep-learned segmentation but requires much smaller training datasets than other deep learning architectures \cite{ronneberger2015u}. Although promising, these methods have not yet been rigorously validated and do not provide continuous PD scores or the corresponding regions of dense tissue segmentation.

Combining conventional image processing methods and machine learning with deep learning techniques can further boost the performance of AI methods in mammographic tasks \cite{kooi2017large}. Here we introduce Deep-LIBRA, an AI method for PD estimation, which combines the U-Net deep learning architecture with image processing and machine learning techniques to segment the breast region, identify the dense tissue areas, and accurately estimate PD. The proposed method was developed using a large multi-ethnic, multi-institution train-validation set totaling 15,661 FFDM images from 4,437 women. Further, it was independently evaluated on 6,478 case-control FFDM images from 1,702 women for assessing both the accuracy of PD estimation and breast cancer risk prediction. Deep-LIBRA has been implemented as open-source software using Python packages and has been made publicly available through \href{https://github.com/CBICA/Deep-LIBRA}{code on GitHub},  \href{https://drive.google.com/drive/folders/1vvnIDYow_YhrRnr8vORxkjorjiHOujOw?usp=sharing}{trained networks on Google Drive},  and \href{https://www.sciencedirect.com/science/article/pii/S1361841521001845}{published paper}.

\section{Methods} \label{methods}
Deep-LIBRA is a pipeline of AI modules sequentially performing all three key computational steps involved in automated PD quantification from FFDM: removal of the image background, removal of the pectoralis muscle, and PD estimation. This section describes the imaging data sets used to develop and evaluate each AI module of Deep-LIBRA.

\begin{table}
  \centering
  \caption{General characteristics of six study datasets. This table shows the screening institute, the number of images and women, range of screening dates, racial distribution, and information about where and how they have been employed in this study.}
\resizebox{0.95\linewidth}{!}{
    \begin{tabular}{l|c c c c c c c}
    \toprule
          & \textbf{DS 1} & \textbf{DS 2} & \textbf{DS 3-a} & \textbf{DS 3-b} & \textbf{DS 4} & \textbf{DS 5} \\
    \midrule
    \textbf{Institute} & HUP   & HUP   & MC  & HUP   & HUP   & MC \\
    \midrule
    \textbf{Number of Images} & 11,200 & 1,100 & 3,314 & 1,147 & 110  & 6,368 \\
    \textbf{Number of Women} & 2,200 & 1,100 & 1,662 & 575   & 110   & 1,592 \\
    \midrule
    \textbf{Screening start date } & 2010  & 2010  & 2008  & 2010  & 2010  & 2013 \\
    \textbf{Screening end date} & 2012  & 2012  & 2012  & 2014  & 2012  & 2015 \\
    \midrule
    \textbf{Caucasians (\% )} & 45    & 45    & 98    & 47    & 45   & 97 \\
    \textbf{African American (\% )} & 45    & 45    & \_    & 53    & 45    & \_ \\
    \textbf{Others (\%) } & 10    & 10    & 2     & \_    & 10    & 3 \\
    \midrule
    \textbf{Used in development} & Yes   & Yes   & Yes   & Yes   & No    & No \\
    \textbf{Cross-validation or Bootstrap} & No    & No    & Yes   & Yes   & No   & Yes \\
    \textbf{Train (\%) } & 90    & 90    & 67    & 67    & \_    & \_ \\
    \textbf{Validation (\%) } & 10    & 10    & 33    & 33    & \_    & \_ \\
    \textbf{Test (\%) } & \_    & \_    & \_    & \_    & 100   & 100 \\
    \midrule
    \textbf{PD accuracy assessment} & No    & No    & Yes   & No   & No   & Yes \\
    \textbf{Case-control classification based on PD} & No    & No    & No    & Yes   & No    & Yes 
    \end{tabular}}
\label{tab_datasets}
\end{table}

\subsection{Study datasets} \label{dataset}
Training, validation, and testing datasets were all collected from retrospective datasets (if breast cancer cases were all priors, negative screenings) of mammographic screening images obtained from two large breast screening practices: the Hospital of the University of Pennsylvania (HUP), Philadelphia, PA, and the Mayo Clinic (MC), Rochester, MN. Our study used raw (i.e., ``FOR PROCESSING'') FFDM images acquired with Selenia or Selenia DimensionsTM units (Hologic Inc, Bedford, MA, USA). These datasets are described in Table \ref{tab_datasets} and Supplementary Materials (Figures 1 and 2).

\subsubsection{Training and validation datasets}
\begin{itemize}
\item \textbf{Dataset to develop the background removal module (ds1):} This dataset consisted of 11,200 images (2,200 women) randomly selected from the HUP screening cohort dataset, and evenly split among left and right lateralities and craniocaudal (CC) and mediolateral oblique (MLO) views \cite{mccarthy2016racial}. This dataset represented the racially diverse screening population at HUP (45\% African American, 45\% Caucasian, and 10\% other ethnicities) \cite{mccarthy2016racial}. 

\item \textbf{Dataset to develop the pectoralis muscle removal module (ds2):} Since the pectoralis muscle is almost always visible only in MLO, the MLO view FFDM images of ds1 were used as the basis of this dataset. Due to the time required for manual delineation of the pectoralis muscle (5 to 10 minutes per FFDM image), 1,100 MLO view images were randomly selected from ds1, maintaining the corresponding racial and laterality distributions.  

\item \textbf{Datasets to develop the PD estimation module (ds3):} One portion of this dataset (ds3-a) was used to guide the development of this module in terms of accuracy in PD estimation, and another (ds3-b) to account for the performance of the PD estimates in breast cancer risk assessment.

\textbf{1.}	Dataset to train and validate the PD estimation module (ds3-a): This consisted of 3,314 FFDM images (1,657 women) from the MC dataset, including the left and right CC views, for which human-rater Cumulus PD values, ``gold-standard'', were available \cite{brandt2016comparison}. This case-control dataset (cases and controls were matched) is described in detail in the Supplementary Materials (Table 1).

\textbf{2.}	Dataset to assess the PD estimation module in breast cancer risk-assessment (ds3-b): We used a previously published case-control dataset (cases and controls were matched) consisting of 1,147 FFDM images from 115 cases and 460 controls acquired at HUP, including left and right MLO views \cite{gastounioti2018incorporating}. This dataset is described in detail in the Supplementary Materials (Table 2). 
\end{itemize}

\begin{figure}[t!]
     \centering
     \includegraphics[width=1\linewidth]{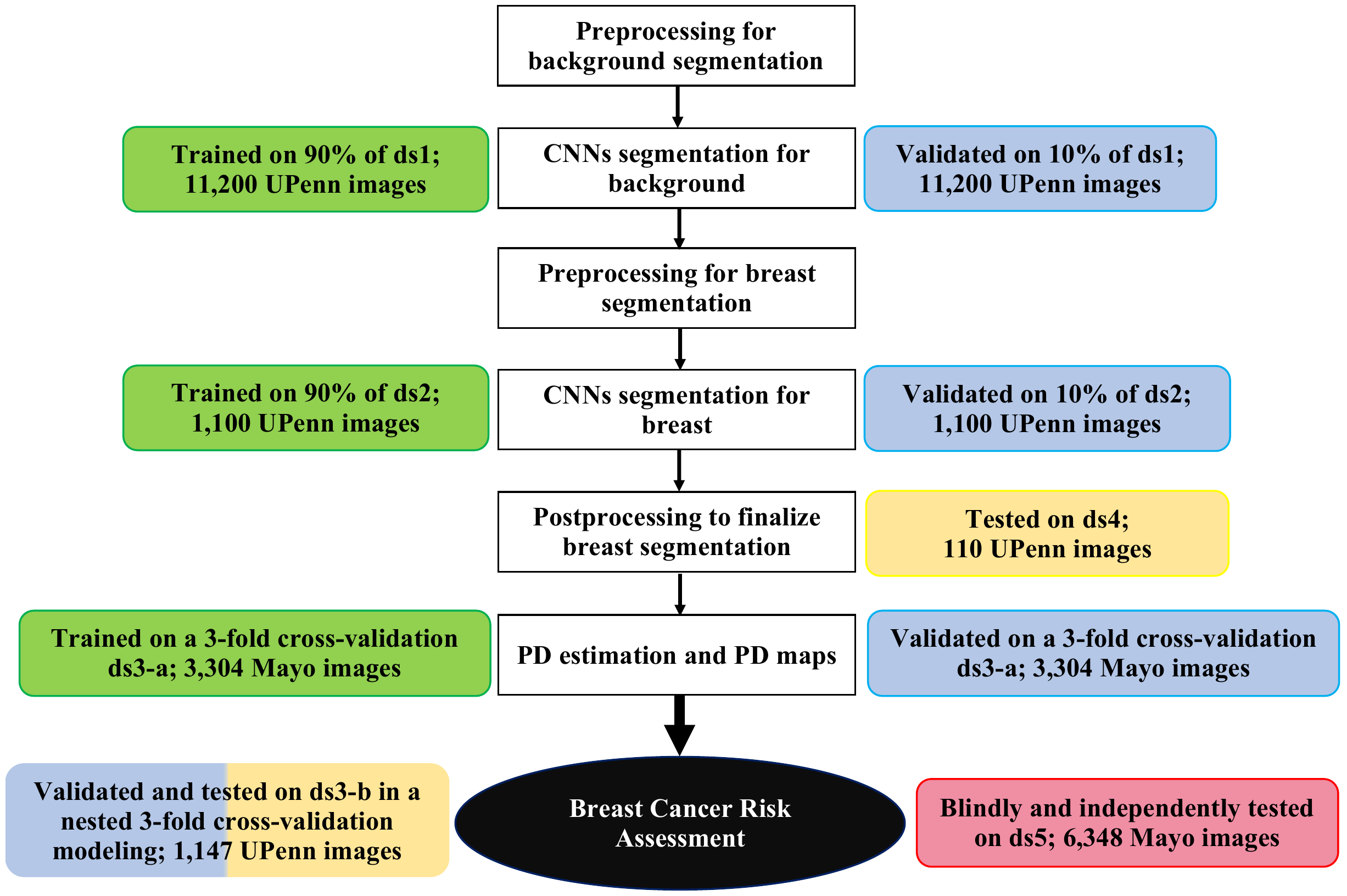} 
		\caption{White boxes: the six steps of the Deep-LIBRA procedure. 
		Green, blue, yellow, and red boxes: training, validation, independent test, and blinded independent test datasets, respectively. 
		UPenn: University of Pennsylvania; Mayo: the Mayo Clinics. 
		Ds3-b was used to validate case-control risk assessment during the PD development phase.} 
\label{fig_data}
\end{figure}

\subsubsection{Independent test datasets}
The following two datasets were used to independently evaluate Deep-LIBRA after development was complete. There was no overlap with images nor women used in training or validation.
\begin{itemize}

\item \textbf{Dataset to evaluate breast segmentation (ds4):} This consisted of 110 randomly selected FFDM MLO view images (110 women) from the HUP screening cohort dataset. The HUP screening cohort dataset included more 50,000 FFDM images \cite{mccarthy2016racial}, and ds1 (ds2 were selected from ds1) and ds3 used a random selection of images from this cohort dataset. The FFDM images were split among left and right lateralities and CC and MLO views. Also, we maintained the corresponding racial and laterality distributions, more details in Table \ref{tab_datasets}.  

\item \textbf{Dataset to evaluate breast cancer risk assessment (ds5):} This consisted of 6,368 FFDM images from MC. Eligible cases with negative screening results were all incident breast cancers reported to the MC tumor registry (n = 414) acquired between 2013 to 2015. Approximately three controls (n = 1,178) without prior breast cancer were matched to each case on age (5-year caliper matching), race, state of residence, FFDM screening exam date, and FFDM screening machine. Mammograms were acquired on median 4.7 [interquartile range (IQR): 4.1, 5.1] years prior to case diagnosis. This dataset included LIBRA area-based PD values, area-based and volumetric Volpara PD values \cite{hartman2008volumetric}, clinical BI-RADS density assessments, and semi-automated Cumulus PD values. This dataset is characterized in detail in the Supplementary Materials (Table 3).

\end{itemize}

\subsection{Algorithm operation} \label{algorithm}
The core of Deep-LIBRA is a sequence of three AI modules for (1) removal of the FFDM image background, (2) removal of the pectoralis muscle, and (3) segmentation of the dense tissue area and subsequent PD estimation. The first two modules are based on deep learning (U-Net); intensity- and texture-based features as input to machine learning methods form the basis of the third module. These modules are outlined in Table \ref{tab_datasets}, Figure \ref{fig_data}, and Supplementary Materials (Figure 1). 

Before applying these modules, standard pre-processing steps for raw FFDM images were applied \cite{keller2012estimation}. Images were log‐transformed, inverted, and squared, and image orientation was standardized.

\begin{figure}[t]
     \centering
     \includegraphics[width=1\linewidth]{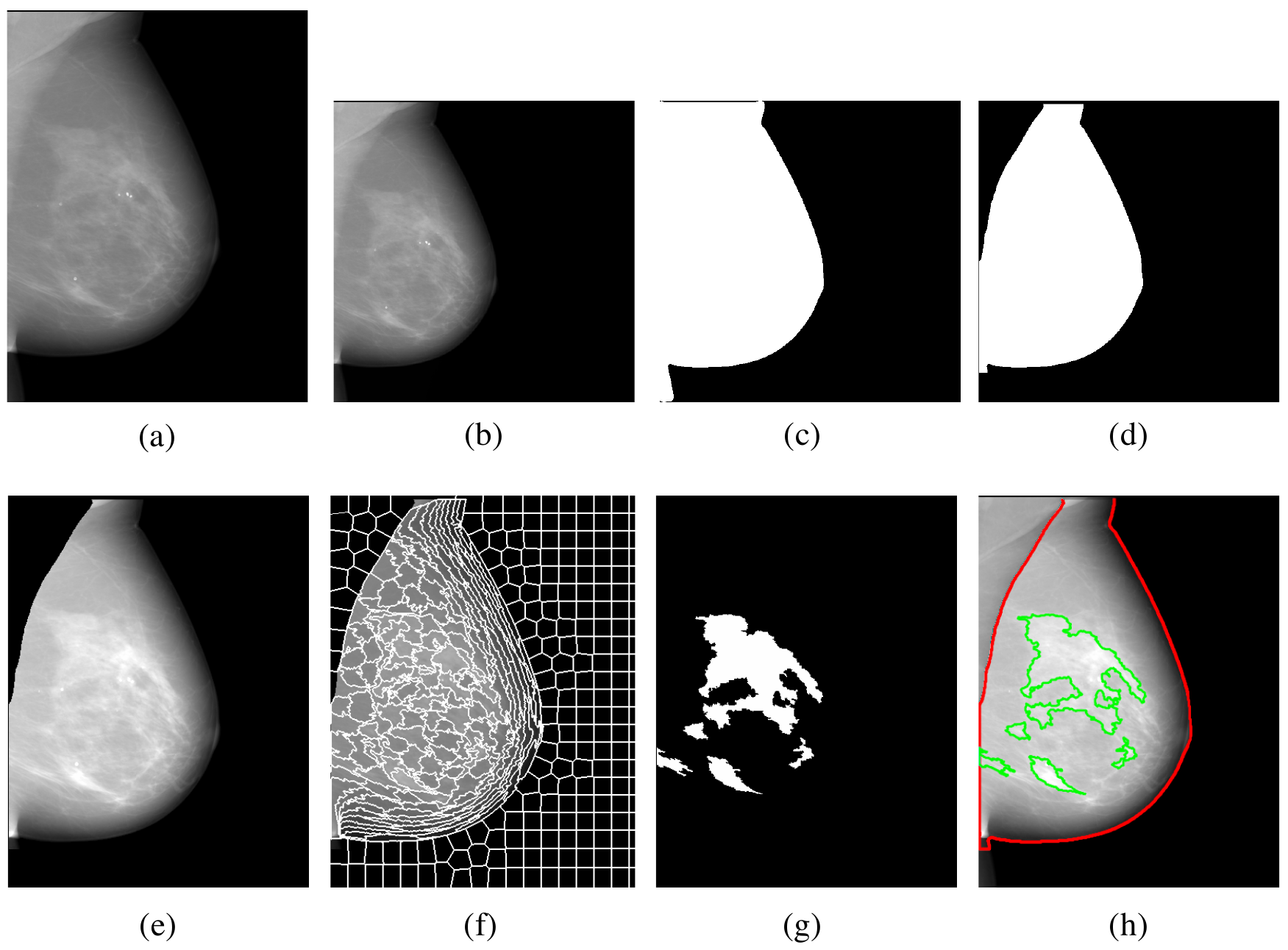} 
		\caption{Detailed illustration of Deep-LIBRA processing in a FFDM image.
Panel (a) shows the original image in 16-bit resolution, and panel (b) is the zero-padded image in an 8-bit intensity resolution.
The zero-padded image is used by the background segmentation CNN, which generates image shown in panel (c).
Panel (d) is the output of pectoralis segmentation using the second CNN following by the breast segmentation shown in panel (e).
The image from panel (e) is used to generate superpixels, panel (f). 
Then, features are extracted from the original image, panel (a).
Finally, SVM classifies the superpixels based on the extracted features, resulting in dense tissue segmentation, as shown in panel (g).
Last but not least, panel (h) shows the dense tissue segmentation overlaid on the original image.} 
\label{fig_steps}
\end{figure}

\subsubsection*{Background removal}
This module performed binary segmentation of the background versus non-background image regions, where the background removed consisted of both air and extraneous objects. Segmentation was implemented via the U-Net architecture, slightly modified by replacing the simple convolutional layers of the encoder with ResNet encoder modules, as these can extract more in-depth information \cite{maghsoudi2020net, szegedy2017inception}. The CNN training was performed using the dataset ds1 and a 90\%–10\% split-sample for training and validation. To further improve U-Net performance, data augmentation \cite{ronneberger2015u} was applied: for each training epoch, each image was randomly altered by any combination of rotation (-22.5 to +22.5 degrees); horizontal shift (-20\% to +20\% of image width);  vertical shift (-20\% to +20\% of image height); zoom (-20\% to +20\%); and horizontal flip. The background region segmented by the U-Net was further refined by removing any regions not connected to any of the four image boundaries. Figure \ref{fig_steps} shows the outcome from this step.

The performance was measured against background segmentation masks generated using the previously validated publicly available software tool, LIBRA \cite{keller2012estimation}. These were further reviewed using ImageJ software \cite{rueden2017imagej2} which resulted in 267 masks being manually revised by OHM. The loss function for training was the inverse weighted Dice measure, defined as 1 - weighted dice \cite{chang2009performance}, to reduce the effect of unbalanced regions in training.

The segmentation performance of the trained CNN was measured on the validation set using four parameters: 1) dice \cite{chang2009performance}, 2) weighted dice, 3) sensitivity \cite{chang2009performance}, and 4) weighted sensitivity. Detailed descriptions of the performance evaluation measures are available in the Supplementary Materials.

\subsubsection{Pectoralis muscle removal}
This module segmented the breast from the image remaining after background removal. As with the background removal module, this was a binary segmentation CNN based on the U-Net architecture. Training used the ds2 dataset, again using a 90\%–10\% split-sample for training and validation, and the inverse weighted dice measure as the loss function. Data augmentation was again applied, with alterations of rotation (-22.5 to +22.5 degrees); horizontal shift (-15\% to +15\% of image width); vertical shift (-15\% to +15\% of image height); and zoom (-15\% to +15\%). The zoom and shift ranges were bounded at 15\%, rather than 20\% used in background removal because the pectoralis occupies an appreciably smaller portion of an image than the background. Any abdominal tissue remaining in the image was removed. The paddle compression effect, a bump of abdominal tissue below the breast caused by paddle compression, was also removed, based on the gradient of the breast contour coordinates (see also Supplementary Materials).

The performance was measured against delineations of the pectoralis muscle manually performed using the ImageJ software by a research scientist (OHM, two years of experience) under the guidance of a fellowship-trained, board-certified, breast imaging radiologist (EFC, more than 25 years of experience). The same four evaluation measures mentioned above were used to evaluate segmentation performance.

\subsubsection{PD calculation and dense versus non-dense image partition}
The PD calculation module involved three major steps: 1) Partitioning the breast into superpixels using image intensity information. These superpixels are what the later steps categorized as dense or non-dense. 2) Calculating texture feature values on the image and on the superpixels. 3) Using these texture features as inputs to machine learning to classify superpixels as either dense or non-dense and thus calculate PD from each FFDM image.

A superpixel is a contiguous subregion of the breast image. By defining superpixels using gray-level intensity values and spatial information, we can generate meaningful localized clusters \cite{achanta2012slic}. To aggregate neighboring pixels into superpixels, we used simple linear iterative clustering (SLIC), a spatially localized version of k-means clustering, which is fast, adheres to local boundaries, and generates superpixels of similar sizes (making the superpixels suitable for representation of scale-variant features such as texture features) \cite{achanta2012slic}. Based on the image size, we partitioned each of all available images into 512 superpixels.

Then, we generated a reference classification of each superpixel as dense versus non-dense, using the image-wise ``gold-standard'' PD scores available for ds3-a. To generate this reference classification, an average intensity is calculated for each superpixel. For a given intensity cutoff, an overall (across all images) PD score is calculated. Similarly, the ``gold-standard'' PD values can be combined into an overall PD. The intensity cutoff that minimizes the difference with the ``gold-standard'' PD values is selected and used to assign each superpixel a reference value of dense versus non-dense region.

The module then computed a total of 101 texture features on each complete image, and those plus an additional 50 features on each superpixel of each image (a list of these features is provided in the Supplementary Materials (Table 6)). Features were extracted using the  PyRadiomics library \cite{van2017computational} and additional Python packages (a full list of packages can be found on \href{https://upenn.box.com/s/9myv3at6cu3kzqsuphv56uwstj0fhoh9}{GitHub}). 

Before using these features in PD modeling, two techniques were used to reduce feature dimensionality. First, for highly-correlated groups of features (absolute Pearson's $r > 0.95$), a single feature from each group was retained (100 features remained from the total of 151 features) which had a maximum interquartile range. Second, a random-forest classifier was applied to all superpixels with the remaining texture features as predictors and the reference dense versus non-dense classification as the supervised classification. From this random forest, the procedure determined the 80 most-predictive features to retain. 

In this module's final step, we trained a support vector machine (SVM) on ds3-a, classifying superpixels as dense versus non-dense with the retained texture features as predictors. Three-fold cross-validation was used, resulting in three trained models. In the subsequent evaluation, each of these models was tested and a combined model assigned dense versus nondense status to superpixels based on a majority vote of the three SVMs (“final model”).

\subsection{Algorithm evaluation} \label{algorithm}
\subsubsection{Evaluation on development datasets}
\begin{itemize}
\item \textbf{Background and pectoralis removal:} Images in the ds1 and ds2 datasets were used to train and validate the U-Net architecture for background and pectoralis muscle removal, respectively. The four performance evaluation measures (dice, weighted dice, sensitivity, and weighted sensitivity) were used to assess Deep-LIBRA's segmentation performance on ds1 and ds2. 

\item \textbf{PD estimation:} Deep-LIBRA training resulted in three SVMs, each trained on two of three folds of ds3-a. For unbiased evaluation, we tested PD estimation by measuring each SVM's performance separately, on only the images that were not in its two training folds. We compared the PD estimation of Deep-LIBRA on each held-out fold versus the PD values given by each of the ``gold-standard'' Cumulus assessments and the LIBRA software on the images in that fold.

\item \textbf{Case/control discrimination:} Using ds3-b, the Deep-LIBRA PD measure's performance was evaluated in the association with breast cancer case/control status  of case/control breast cancer status. For each of five density measures — Deep-LIBRA, LIBRA, volumetric Quantra (V-Quantra), area-based Quantra (A-Quantra), and categorical BI-RADS — case/control status was modeled as the outcome in each of two logistic regression models: an adjusted model (just for the “final model”), consisting of age, body-mass index (BMI), and the density measure; and an unadjusted model (for the three SVMs and “final model”) consisting of the density measure alone. Model discriminatory ability was assessed via mean area under curve (AUC) of receiver operation curve (ROC) across 100 bootstrap samples (bootstrap samples maintained case-control matching), with CIs derived from those 100 repetitions. Additionally, we examined differences in the PD scores for cases versus controls using the Wilcoxon rank sum test.
\end{itemize}

\subsubsection{Evaluation on independent testing datasets}
\begin{itemize}
\item \textbf{Breast segmentation:} Images in the ds4 dataset were processed to evaluate the final segmented breast area on an independent dataset. The Dice measure (explained in the Supplementary Materials) was calculated, and we compared the Dice values based on Deep-LIBRA to those obtained from LIBRA using a two-sided t-test.

\item \textbf{Discrimination of case-control status:} The association of Deep-LIBRA with breast cancer status on ds5, unlike any other assessments, was performed by an independent analyst at MC, and the algorithm developers were blinded to case-control status. 

For each of the six density measures — Deep-LIBRA, LIBRA, volumetric Volpara, area-based Volpara, Cumulus (the gold standard), and categorical BI-RADS — case/control status was modeled as the outcome in an adjusted conditional logistic regression model \cite{breslow1978estimation}, consisting of age, BMI, and the density measure. Model discriminatory ability was assessed via AUC and effect sizes as odds ratio (OR) per one standard deviation of the model's PD. Additionally, we tested PD values for differences between cases and controls using the Mann-Whitney test. Before using them in modeling, PD measures were transformed to achieve normality: natural logarithm (LIBRA and volumetric Volpara), square root (Cumulus and Deep-LIBRA), or no transform (area-based Volpara). AUCs and their CIs were calculated using a procedure accounting for the matched data and the conditional logistic regression.  ORs were also calculated with the standard CIs. P values for both AUCs and ORs, versus the null hypothesis of no difference from the AUC or OR derived from Deep-LIBRA, were estimated from testing across 1,000 bootstrap replicates (bootstrap samples maintained case-control matching). Finally, we also examined all the PD measures' distributions and compared their correlations using the Spearman correlation coefficient. SAS version 9.4 (Cary, NC) was used for all statistical analyses, and p values considered statistically significant at the 0.05 cutoff.
\end{itemize}

\begin{figure}[t]
     \centering
     \includegraphics[width=1\linewidth]{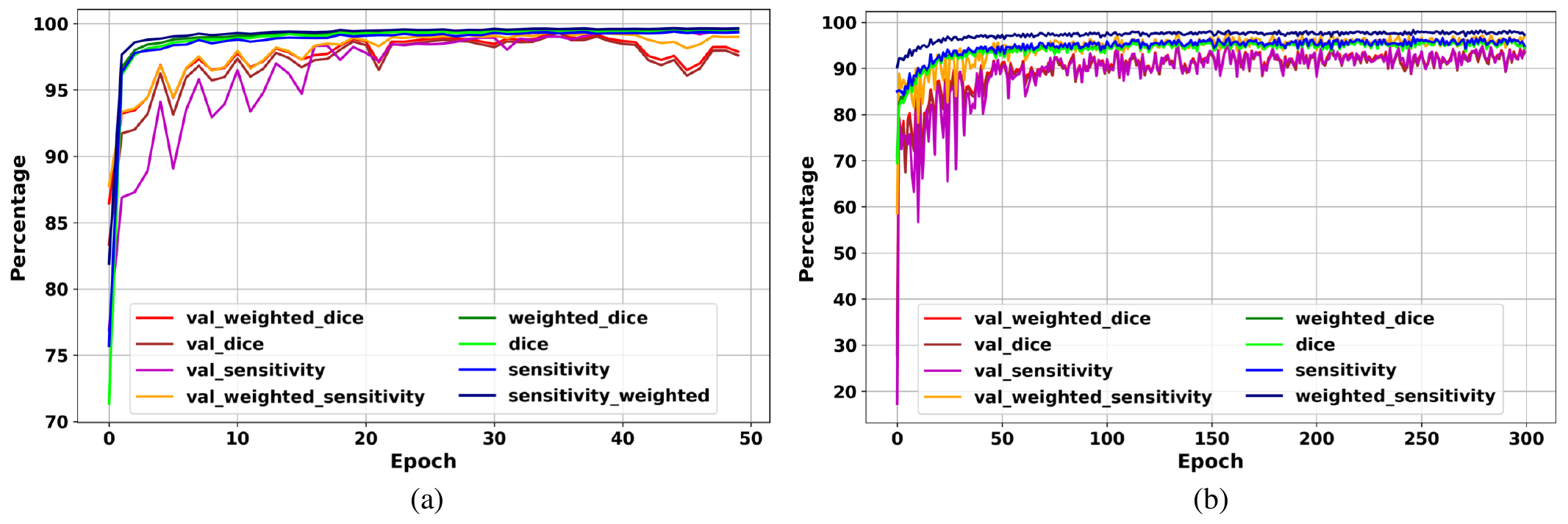} 
		\caption{Evaluation curves during the development phase.
Panels (a) and (b) show the training and validation results for background and breast segmentation U-Nets, respectively.
Four measures are calculated for each epoch of training.
Validation results are proceeded as ``val\_." 
As the panel (b) shows, there is no sign of overfitting for pectoralis muscle segmentation while panel (b) indicates some possible signs of overfitting after epoch 40 shown by a wider fluctuation range in results on the validation set (a drop of 5\% in the dice score).} 
\label{fig_step1}
\end{figure}

\section{Results}
\subsection{Performance measures and evaluation on development datasets}
\subsubsection{Background and pectoralis removal}
Both the background removal and pectoralis removal modules achieved high performance and were sufficiently trained. The highest weighted dice score achieved by background removal on the validation set was 99.4\% after 35 epochs, and it achieved 99.5\% on the training set at the same epoch (Figure \ref{fig_step1}). Pectoralis removal achieved the highest weighted dice of 95.0\% on the validation set after 158 epochs, and 96.3\% on the training set at the same epoch (Figure \ref{fig_step1}).

\subsubsection{PD estimation}
Deep-LIBRA PD estimation on ds3-a showed good agreement with ``gold-standard'' PD values (Figure \ref{fig_step2}). Mean PD differences between Deep-LIBRA and Cumulus, for each of the three Deep-LIBRA trained SVMs (each evaluated only on its testing fold), were 4.91 [95\% CI: 4.48, 5.34], 4.64 [95\% CI: 4.31, 4.99], and 4.22 [95\% CI: 3.95, 4.49]; mean differences of LIBRA PD from Cumulus, on those folds, were 5.28 [95\% CI: 4.95, 5.60], 5.24 [95\% CI: 4.96, 5.52], and 5.39 [95\% CI: 5.08, 5.70]. For two of the three folds, differences between Deep-LIBRA PD values and those of Cumulus were statistically significantly less than those between the LIBRA values and Cumulus using a paired two-sided t-test (p values 0.179, 0.008, and 0.001). As well, Spearman correlations between Deep-LIBRA and Cumulus PD (0.80, 0.79, and 0.84) were higher than those between LIBRA and Cumulus (0.70, 0.70, and 0.69) for all three folds. 

\subsubsection{Case/control discrimination}
\begin{figure}[!t]
     \centering
     \includegraphics[width=1\linewidth]{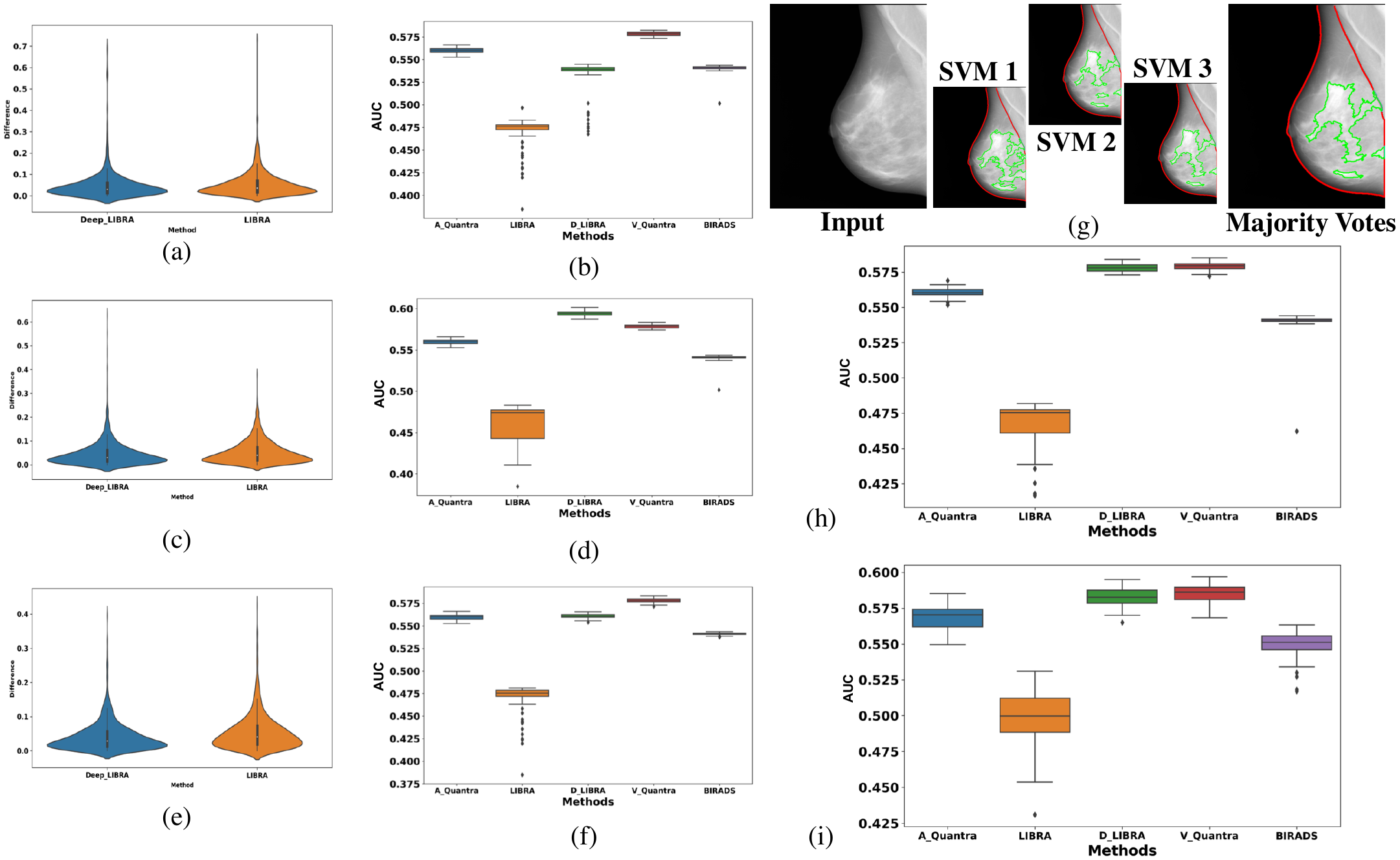} 
		\caption{PD comparisons for three-fold cross validation on ds3-a and majority voting approach.
Panels (a), (c), and (e) show the absolute difference of PD values between Cumulus (reference) and Deep-LIBRA or LIBRA calculated on the three folds of ds3-a.
Panels (b), (d), and (f): AUC of case/control unadjusted logistic regression using each  density measure: area-based Quantra, LIBRA, Deep-LIBRA, volumetric Quantra, and ordinal BI-RADS.
Deep-LIBRA has the best agreement (less average for the absolute differences) for all three folds, and its ranking varies from first to third.
Panel (g) using three SVM models to make the final PD estimation using the majority vote for each superpixel. 
Panels (h) and (i) show AUC of case/control adjusted logistic regression (model covariates age, BMI, and density measure) using each  density measure: area-based Quantra, LIBRA, Deep-LIBRA, volumetric Quantra, and ordinal BI-RADS.} 
\label{fig_step2}
\end{figure}

In the task of logistic regression of case/control status on ds3-b, in the unadjusted model, the three Deep-LIBRA SVMs (each evaluated on its test fold) yielded mean AUCs of 0.532, 0.594, and 0.561. Results using the other density measures tested were similar: V-Quantra 0.578, A-Quantra 0.560, LIBRA 0.469, and ordinal BI-RADS 0.541 (Table \ref{tab_casecontrol}, Figure \ref{fig_step2}).

We also evaluated a combined measure, in which each superpixel's dense/non-dense classification was assigned as the majority vote of the three SVMs, and these assignments were used in the resulting PD estimation and case/control modeling. This was motivated by the varying results we observed among the three SVM models and allowed Deep-LIBRA to reduce data partitioning's random effect. The concept of majority voting on a sample FFDM is illustrated in Figure \ref{fig_step2}.

Evaluating this combined measure on ds3-b in the unadjusted logistic regression model gave a mean AUC of 0.578, equal to that from V-Quantra and greater than the AUCs of the other density measures. In the adjusted logistic regression model, the combined measure gave an AUC of 0.582 for Deep-LIBRA — close to that for V-Quantra, and greater than those for other density measures.

\subsection{Performance measures and evaluation on independent testing datasets}
The following assessments were performed after finalizing the algorithm on datasets unknown to algorithm development.

\subsubsection{Breast segmentation}
Deep-LIBRA gave a mean dice score of 92.5\%, which was statistically significantly better (p <0.001) than LIBRA (mean dice 83.4\%) (details in Supplementary Materials, Table 4). LIBRA tends to consider a line for perctoralis muscle which can lead to an inaccurate results. This point is clarified in Supplementary Materials (Figure 4) by showing the segmented regions of five FFDM images from LIBRA and Deep-LIBRA. 

\subsubsection{PD distributions and correlation}
Cumulus (calculated on CC view images, and averaged across lateralities) had a median PD value of 12.4 [IQR: 6.6, 12.6] and 15.5 [IQR: 8.8, 25.3] for controls and cases, respectively. Deep-LIBRA averaged over four views (two CC and two MLO) gave median values of 11.5 [IQR: 5.7, 19.9] and 14.1 [IQR: 7.7, 23.9] for controls and cases. LIBRA averaged over four views gave 10.8 [IQR: 8.3, 16.0] and 11.9 [IQR: 8.3, 18.9], and volumetric-based Volpara 6.0 [IQR: 4.4, 9.9] and 6.7 [IQR: 4.6, 11.7], again, for controls and cases respectively. Area-based Volpara PD values were much higher than those from other techniques, with median PD of 48.0 [IQR: 28.9, 68.3] for controls and 53.8 [IQR: 32.9, 76.7] for cases. The four BI-RADS classes — almost entirely fat tissue, scattered fibroglandular tissue, heterogeneous fibroglandular tissue, and extreme fibroglandular tissue — had category percentages of 22\%, 41\%, 31\%, and 6\% for controls and 15\%, 36\%, 41\%, and 8\% for cases. All six measures showed higher density for cases versus controls with statistically significant differences (complete details, Table \ref{tab_PD_mayo}).

Deep-LIBRA PD had Spearman correlation coefficients of 0.90, 0.89, and 0.83 with Cumulus PD for values averaged across four views, CC views, and MLO views, respectively. Area-based and volumetric-based Volpara, from four views, gave correlations of 0.89 and 0.91 with Cumulus, respectively, and LIBRA from four views gave 0.76. (See the Supplementary Materials (Figure 3) for more on the associations between all PD measures.)

Results from the Deep-LIBRA PDs derived from averaging over each of four views, CC views only, and MLO views only yielded mean AUCs of 0.612 [95\% CI: 0.583, 0.640], 0.611 [95\% CI: 0.583, 0.639], and 0.596 [95\% CI: 0.568, 0.624], respectively, as compared to a Cumulus value (provided  for CC view) of 0.619 [95\% CI: 0.592, 0.647], with p = 0.85 for difference between Cumulus and the corresponding CC-view Deep-LIBRA AUCs. Volumetric Volpara (four-view) gave a mean AUC of 0.599 [95\% CI: 0.572, 0.627; p = 0.37]; area-based Volpara (four-view) gave 0.578 [95\% CI: 0.551, 0.607; p = 0.04]; LIBRA (four-view) gave 0.564 [95\% CI: 0.535, 0.592; p = 0.01; and BI-RADS gave 0.596 [95\% CI: 0.568, 0.624; p = 0.35].

ORs for the association of density with breast cancer for four-view and CC-view Deep-LIBRA, Cumulus, MLO-view volumetric Volpara, and MLO-view area-based Volpara all had similarly high values: 1.61 [95\% CI: 1.37, 1.88]) and 1.64 [95\% CI: 1.40, 1.91] for Deep-LIBRA; Cumulus 1.64 [95\% CI: 1.39, 1.93]; Volpara 1.62 [95\% CI: 1.37, 1.92] and 1.62 [95\% CI: 1.28, 1.79]. Other values were lower (Table \ref{tab_cc_mayo}).

\begin{table}[t]
  \centering
  \caption{AUC of discrimination of case-control status via logistic regression, using Deep-LIBRA and other PD measures in the development phase (dataset ds3-b). 
  A-Quantra and V-Quantra: area-based Quantra and volumetric-based Quantra PD values. 
 Each of folds 1, 2, and 3 is the result of one of the Deep-LIBRA SVMs on its testing data from ds3-b . 
 Final UA and Final A are the results of unadjusted and adjusted modeling. }
  \resizebox{1.0\linewidth}{!}{
    \begin{tabular}{l| c c c c c}
    \toprule
          & \textbf{LIBRA} & \textbf{V\_Quantra} & \textbf{A\_Quantra} & \textbf{BIRADS} & \textbf{Deep\_LIBRA} \\
    \midrule
    \textbf{Fold 1 UA} & 0.469 & 0.578  & 0.560 & 0.541 & 0.532 \\
    \textbf{Fold 2 UA} & 0.460 & 0.579 & 0.560 & 0.541 & 0.594 \\
    \textbf{Fold 3 UA} & 0.467 & 0.578 & 0.561 & 0.541 & 0.561 \\
    \midrule
    \textbf{Final UA} & 0.467 & 0.579 & 0.561 & 0.540 & 0.578 \\
    \textbf{Final A} & 0.498 & 0.586 & 0.568 & 0.550 & 0.582 
    \end{tabular}}
\label{tab_casecontrol}
\end{table}

\section{Discussion}
Deep-LIBRA is an open-source AI-based tool for fully automated calculation of PD from FFDM. Accurate evaluation of PD is crucial in clinical practice, as density is both a risk factor for breast cancer and decreases the information available from FFDM images. Knowledge of PD influences recommendations regarding risk, indications for further imaging, or assessment of changes. Deep-LIBRA can provide breast segmentation by separating the dense from the non-dense tissues within the breast region. The method included two major steps: the segmentation of the breast region and the estimation of PD. It employed some pre-processing steps, two CNNs (the modified U-Net architecture), superpixels, extracting texture features, and SVM to accomplish the PD estimation.

We employed two binary segmentation U-Nets, one for background and one for pectoral removal instead of one multi-class network. This was done because we first wanted to reduce some variations in the input images (e.g., background intensity range and pattern can significantly vary in images). Second, we wanted to remove unpredictable artifacts in some cases, such as paddles, rings, and other materials. As shown in Figure \ref{fig_step1}, the segmentation results did not depict an obvious sign of overfitting. One of the reasons could be the use of the two U-Nets. The breast segmentation results on ds4 showed that our proposed method was significantly superior compared to LIBRA (p-value <0.001). 

For developing the PD module, we initially considered three-fold cross-validation on ds3-a to train three SVMs. The Cumulus score's absolute differences with the Deep-LIBRA and LIBRA PD scores were marginally different, but the Deep-LIBRA PD scores had the lowest difference in comparison with Cumulus PD scores. Simultaneously, each of the three SVMs was separately tested on ds3-a to evaluate the model's performance in breast cancer risk assessment. The results, illustrated in Figure \ref{fig_step2}, varied by selecting folds for training and validation sets. To overcome these variations, we used the majority voting approach (including all three folds) to minimize the chance of such variations affecting our final model.

\begin{table}[t]
 \centering
 \caption{The PD measures on ds5.
 Cumulus, Deep-LIBRA, LIBRA, and Volpara measures are median values with interquartile ranges in parentheses. 
  BIRADS data shows the number of women in each BIRADS density category with the associated percentage relative to the total number of women.
 Cumulus measures are estimated on the CC view and based on the processed images, there are considered as the gold standard measure for PD.
 Deep-LIBRA has the closest mean and the most similar interquartile range with Cumulus.}
\resizebox{0.9\linewidth}{!}{
  \begin{tabular}{ l| c c c}
  \toprule
  \textbf{Breast density measure} & \textbf{Controls (N=1178)} & \textbf{Cases (N=414)} & \textbf{P Value} \\
  \midrule
  \textbf{Deep-LIBRA PD (4 views), median (Q1, Q3)} & 11.5 (5.7, 19.9) & 14.1 (7.7, 23.9) & \textless.001 \\
  \textbf{Deep-LIBRA PD (CC), median (Q1, Q3)} & 12.0 (6.6, 20.6) & 15.7 (8.6, 25.7) & \textless.001\\
  \textbf{Deep-LIBRA PD (MLO), median (Q1, Q3)} & 10.3 (4.4, 19.8) & 13.4 (6.1, 22.1) & \textless.001 \\
   \midrule
  \textbf{Cumulus Proc PD (CC), median (Q1, Q3)} & 12.4 (6.6, 21.6) & 15.5 (8.8, 25.3) & \textless.001 \\
  \midrule
  \textbf{LIBRA PD (4 views), median (Q1, Q3) } & 10.8 (7.6, 16.0) & 11.9 (8.3, 18.9) & 0.004 \\
  \textbf{LIBRA PD (CC), median (Q1, Q3)} & 10.5 (7.1, 15.6) & 11.5 (7.6, 18.7) & 0.006 \\
  \textbf{LIBRA PD (MLO), median (Q1, Q3)} & 11.2 (7.6, 16.9) & 12.3 (8.0, 19.4) & 0.005 \\
  \midrule
  \textbf{Volumetric Volpara PD (4 views), median (Q1, Q3)} & 6.0 (4.4, 10.0) & 6.7 (4.6, 12.0) & \textless.001 \\
  \textbf{Volumetric Volpara PD (CC), median (Q1, Q3)} & 6.0 (4.4, 9.9) & 6.7 (4.6, 11.7) & \textless.001 \\
  \textbf{Volumetric Volpara PD (MLO), median (Q1, Q3)} & 5.9 (4.3, 9.7) & 6.8 (4.7, 12.7) & \textless.001 \\
  \midrule
  \textbf{Area Volpara PD (4 views), median (Q1, Q3)} & 48 (28.9, 68.3) & 53.8 (32.9, 76.7) & \textless.001 \\
  \midrule
  \textbf{BIRADS Density}, n (\%) & & & \textless.001\\
  \textbf{.   1} & 255 (22\%) & 61 (15\%) & \\
  \textbf{.   2} & 487 (41\%) & 149 (36\%) & \\
  \textbf{.   3} & 364 (31\%) & 170 (41\%) & \\
  \textbf{.   4} & 71 (6\%) & 34 (8\%) &
  \end{tabular}}
\label{tab_PD_mayo}
\end{table}

Analysis of ds5 could shed more light on various aspects of the Deep-LIBRA method and  may offer the most important assessment. First, the developer remained blinded to the ds5 measures and methods for analysis; second, ds5 is an independent test set; third, ds5 is a large case-control dataset with available Cumulus, Volpara, LIBRA, and BI-RADS PD scores; and fourth, the most important role of assessing PD is to predict who will get breast cancer. The first assessment was a comparison of PD estimation between Deep-LIBRA, Volpara (area-based and volumetric-based measures), LIBRA, Cumulus, BI-RADS, and the ``gold-standard'' human-dependent values. Deep-LIBRA PD scores were the most similar to Cumulus for both cases and controls, as illustrated in Table \ref{tab_PD_mayo}. 

The second assessment on ds5 was based on the Spearman correlation. The results, summarized in the Supplementary Materials (Table 5 and Figure 3), showed that the correlation of PD scores between Deep-LIBRA and Cumulus were above 0.90 for four views and CC views while it was 0.83 for MLO views.

\begin{table}[t]
 \centering
 \caption{Association of Density Measures with Breast Cancer on ds5. 
AUCs and CIs  calculated on whole datasets. 
AUC CIs calculated accounting for matched nature of conditional logistic regression. 
P values for both AUCs and ORs obtained from testing across 1,000 bootstrap replicates, versus  null hypothesis of no difference from the AUC or OR derived from Deep-LIBRA using the same views. }
 \resizebox{1.0\linewidth}{!}{
  \begin{tabular}{ l | c  c  c  c}
  \toprule
  \textbf{Density score} & \textbf{OR (95\% CI)} & \textbf{p-value} & \textbf{AUC (95\% CI)} & \textbf{p-value} \\
  \midrule
  \textbf{Deep-LIBRA PD (4 views)} & 1.61 (1.37, 1.88) & -- & 0.612 (0.584, 0.640) & -- \\
  \textbf{Deep-LIBRA PD (CC)} & 1.64 (1.40, 1.91) & -- & 0.611 (0.583, 0.639) & -- \\
  \textbf{Deep-LIBRA PD (MLO)} & 1.46 (1.26, 1.69) & -- & 0.596 (0.568, 0.624) & -- \\
  \midrule
  \textbf{Cumulus PD (CC)} & 1.64 (1.39, 1.93) & 0.99 & 0.619 (0.592, 0.647) & 0.85\\
  \midrule
  \textbf{LIBRA PD (4 views)} & 1.26 (1.09, 1.46) & \textless.001 & 0.564 (0.535, 0.592) & 0.01\\
  \textbf{LIBRA PD (CC)} & 1.19 (1.04, 1.36) & \textless.001 & 0.557 (0.528, 0.585) & 0.01\\
  \textbf{LIBRA PD (MLO)} & 1.26 (1.10, 1.46) & 0.07 & 0.561 (0.533, 0.589) & 0.04\\
  \midrule
  \textbf{Volumetric Volpara PD (4 views)} & 1.55 (1.31, 1.82) & 0.43 & 0.599 (0.572, 0.627) & 0.37\\
  \textbf{Volumetric Volpara PD (CC)} & 1.45 (1.24, 1.71) & 0.02 & 0.588 (0.559, 0.616) & 0.09\\
  \textbf{Volumetric Volpara PD (MLO)} & 1.62 (1.37, 1.92) & 0.10 & 0.598 (0.570, 0.626) & 0.88\\
  \midrule
  \textbf{Area Volpara PD (4 views)} & 1.48 (1.25, 1.74) & 0.10 & 0.578 (0.551, 0.607) & 0.04\\
  \textbf{Area Volpara PD (CC)} & 1.38 (1.18, 1.61) & \textless.001 & 0.567 (0.539, 0.596) & 0.01\\
  \textbf{Area Volpara PD (MLO)} & 1.62 (1.28, 1.79) & 0.53 & 0.591 (0.563, 0.619) & 0.49\\
  \midrule
  \textbf{BI-RADS density} & 1.54 (1.30, 1.81) & 0.45 & 0.596 (0.568, 0.624) & 0.35
  \end{tabular}}
\label{tab_cc_mayo}
\end{table}

The final assessment was to evaluate association of density measures with case-control status using the logistic regression model. ORs confirmed that Deep-LIBRA PD scores had the best agreement with the gold standard compared to other PD scores. In the clinically-motivated test of breast cancer case/control discrimination on an independent test set, the final regression model adjusted for the covariates age, BMI, and Deep-LIBRA PD, achieved an AUC of 0.61. This was comparable to the 0.62 achieved by the model including human-dependent Cumulus PD and superior to that achieved by the other density measures tested. 

Deep-LIBRA's performance suggests the effectiveness of combining different learning technologies and information sources for PD subtasks: Deep-LIBRA employs deep learning on image intensity for background and pectoralis removal, and machine learning (SVMs) on both intensity and texture features for dense/non-dense segmentation. Development revealed a sensitivity of PD calculation to the randomness of the data partition, but we found that PD calculation based on a majority rule among the trained SVMs removed this weakness.

The limitations of this study must also be noted. At this point, Deep-LIBRA has only been trained on ``FOR PROCESSING'' raw FFDM images and one manufacturer (Hologic). Motivated by the Deep-LIBRA performance reported here compared to the other measures, we plan to address these two limitations in our future work by further training and reevaluating Deep-LIBRA for ``FOR PRESENTATION'' vendor-processed FFDM images, while utilizing multi-vendor datasets. Another limitation of this study was using the assessed ``gold standard'' Cumulus PD score by one expert, for which we will also investigate the effect of multiple readers in our future studies. We will consider training Deep-LIBRA using Volpara to assess volumetric PD scores too. Also, further developments in AI could help us improve our method performance and explore new horizons in this field. Finally, as publicly-available open-source software, we hope that Deep-LIBRA will benefit from widespread adoption and collaborative improvements.

\section*{References}
\bibliographystyle{naturemag}
\bibliography{References}

\begin{thebibliography}{50}
\expandafter\ifx\csname natexlab\endcsname\relax\def\natexlab#1{#1}\fi
\providecommand{\url}[1]{\texttt{#1}}
\providecommand{\href}[2]{#2}
\providecommand{\path}[1]{#1}
\providecommand{\DOIprefix}{doi:}
\providecommand{\ArXivprefix}{arXiv:}
\providecommand{\URLprefix}{URL: }
\providecommand{\Pubmedprefix}{pmid:}
\providecommand{\doi}[1]{\href{http://dx.doi.org/#1}{\path{#1}}}
\providecommand{\Pubmed}[1]{\href{pmid:#1}{\path{#1}}}
\providecommand{\bibinfo}[2]{#2}
\ifx\xfnm\relax \def\xfnm[#1]{\unskip,\space#1}\fi
\bibitem[{iCA()}]{iCAD}
, .
\newblock \bibinfo{howpublished}{\url{http://www.icadmed.com}}.
\newblock \bibinfo{note}{Accessed: 2020-03-10}.
\bibitem[{Achanta et~al.(2012)Achanta, Shaji, Smith, Lucchi, Fua and
  S{\"u}sstrunk}]{achanta2012slic}
\bibinfo{author}{Achanta, R.}, \bibinfo{author}{Shaji, A.},
  \bibinfo{author}{Smith, K.}, \bibinfo{author}{Lucchi, A.},
  \bibinfo{author}{Fua, P.}, \bibinfo{author}{S{\"u}sstrunk, S.},
  \bibinfo{year}{2012}.
\newblock \bibinfo{title}{Slic superpixels compared to state-of-the-art
  superpixel methods}.
\newblock \bibinfo{journal}{IEEE transactions on pattern analysis and machine
  intelligence} \bibinfo{volume}{34}, \bibinfo{pages}{2274--2282}.
\bibitem[{Anitha et~al.(2017)Anitha, Peter and Pandian}]{anitha2017dual}
\bibinfo{author}{Anitha, J.}, \bibinfo{author}{Peter, J.D.},
  \bibinfo{author}{Pandian, S.I.A.}, \bibinfo{year}{2017}.
\newblock \bibinfo{title}{A dual stage adaptive thresholding (dusat) for
  automatic mass detection in mammograms}.
\newblock \bibinfo{journal}{Computer methods and programs in biomedicine}
  \bibinfo{volume}{138}, \bibinfo{pages}{93--104}.
\bibitem[{Becker et~al.(2017)Becker, Marcon, Ghafoor, Wurnig, Frauenfelder and
  Boss}]{becker2017deep}
\bibinfo{author}{Becker, A.S.}, \bibinfo{author}{Marcon, M.},
  \bibinfo{author}{Ghafoor, S.}, \bibinfo{author}{Wurnig, M.C.},
  \bibinfo{author}{Frauenfelder, T.}, \bibinfo{author}{Boss, A.},
  \bibinfo{year}{2017}.
\newblock \bibinfo{title}{Deep learning in mammography: diagnostic accuracy of
  a multipurpose image analysis software in the detection of breast cancer}.
\newblock \bibinfo{journal}{Investigative radiology} \bibinfo{volume}{52},
  \bibinfo{pages}{434--440}.
\bibitem[{Brandt et~al.(2016)Brandt, Scott, Ma, Mahmoudzadeh, Jensen, Whaley,
  Wu, Malkov, Hruska, Norman et~al.}]{brandt2016comparison}
\bibinfo{author}{Brandt, K.R.}, \bibinfo{author}{Scott, C.G.},
  \bibinfo{author}{Ma, L.}, \bibinfo{author}{Mahmoudzadeh, A.P.},
  \bibinfo{author}{Jensen, M.R.}, \bibinfo{author}{Whaley, D.H.},
  \bibinfo{author}{Wu, F.F.}, \bibinfo{author}{Malkov, S.},
  \bibinfo{author}{Hruska, C.B.}, \bibinfo{author}{Norman, A.D.}, et~al.,
  \bibinfo{year}{2016}.
\newblock \bibinfo{title}{Comparison of clinical and automated breast density
  measurements: implications for risk prediction and supplemental screening}.
\newblock \bibinfo{journal}{Radiology} \bibinfo{volume}{279},
  \bibinfo{pages}{710--719}.
\bibitem[{Brentnall et~al.(2018)Brentnall, Cuzick, Buist and
  Bowles}]{brentnall2018long}
\bibinfo{author}{Brentnall, A.R.}, \bibinfo{author}{Cuzick, J.},
  \bibinfo{author}{Buist, D.S.}, \bibinfo{author}{Bowles, E.J.A.},
  \bibinfo{year}{2018}.
\newblock \bibinfo{title}{Long-term accuracy of breast cancer risk assessment
  combining classic risk factors and breast density}.
\newblock \bibinfo{journal}{JAMA oncology} \bibinfo{volume}{4},
  \bibinfo{pages}{e180174--e180174}.
\bibitem[{Breslow et~al.(1978)Breslow, Day, Halvorsen, Prentice and
  Sabai}]{breslow1978estimation}
\bibinfo{author}{Breslow, N.}, \bibinfo{author}{Day, N.},
  \bibinfo{author}{Halvorsen, K.}, \bibinfo{author}{Prentice, R.},
  \bibinfo{author}{Sabai, C.}, \bibinfo{year}{1978}.
\newblock \bibinfo{title}{Estimation of multiple relative risk functions in
  matched case-control studies}.
\newblock \bibinfo{journal}{American Journal of Epidemiology}
  \bibinfo{volume}{108}, \bibinfo{pages}{299--307}.
\bibitem[{Chang et~al.(2009)Chang, Zhuang, Valentino and
  Chu}]{chang2009performance}
\bibinfo{author}{Chang, H.H.}, \bibinfo{author}{Zhuang, A.H.},
  \bibinfo{author}{Valentino, D.J.}, \bibinfo{author}{Chu, W.C.},
  \bibinfo{year}{2009}.
\newblock \bibinfo{title}{Performance measure characterization for evaluating
  neuroimage segmentation algorithms}.
\newblock \bibinfo{journal}{Neuroimage} \bibinfo{volume}{47},
  \bibinfo{pages}{122--135}.
\bibitem[{Czaplicka and W{\l}odarczyk(2011)}]{czaplicka2011automatic}
\bibinfo{author}{Czaplicka, K.}, \bibinfo{author}{W{\l}odarczyk, J.},
  \bibinfo{year}{2011}.
\newblock \bibinfo{title}{Automatic breast-line and pectoral muscle
  segmentation}.
\newblock \bibinfo{journal}{Schedae Informaticae} \bibinfo{volume}{20}.
\bibitem[{Dalm{\i}{\c{s}} et~al.(2017)Dalm{\i}{\c{s}}, Litjens, Holland, Setio,
  Mann, Karssemeijer and Gubern-M{\'e}rida}]{dalmics2017using}
\bibinfo{author}{Dalm{\i}{\c{s}}, M.U.}, \bibinfo{author}{Litjens, G.},
  \bibinfo{author}{Holland, K.}, \bibinfo{author}{Setio, A.},
  \bibinfo{author}{Mann, R.}, \bibinfo{author}{Karssemeijer, N.},
  \bibinfo{author}{Gubern-M{\'e}rida, A.}, \bibinfo{year}{2017}.
\newblock \bibinfo{title}{Using deep learning to segment breast and
  fibroglandular tissue in mri volumes}.
\newblock \bibinfo{journal}{Medical physics} \bibinfo{volume}{44},
  \bibinfo{pages}{533--546}.
\bibitem[{D’orsi et~al.(2003)D’orsi, Bassett, Berg, Feig, Jackson, Kopans
  et~al.}]{d2003breast}
\bibinfo{author}{D’orsi, C.}, \bibinfo{author}{Bassett, L.},
  \bibinfo{author}{Berg, W.}, \bibinfo{author}{Feig, S.},
  \bibinfo{author}{Jackson, V.}, \bibinfo{author}{Kopans, D.}, et~al.,
  \bibinfo{year}{2003}.
\newblock \bibinfo{title}{Breast imaging reporting and data system: Acr
  bi-rads-mammography}.
\newblock \bibinfo{journal}{American College of Radiology (ACR), Reston} ,
  \bibinfo{pages}{230--234}.
\bibitem[{Engmann et~al.(2017)Engmann, Golmakani, Miglioretti, Sprague and
  Kerlikowske}]{engmann2017population}
\bibinfo{author}{Engmann, N.J.}, \bibinfo{author}{Golmakani, M.K.},
  \bibinfo{author}{Miglioretti, D.L.}, \bibinfo{author}{Sprague, B.L.},
  \bibinfo{author}{Kerlikowske, K.}, \bibinfo{year}{2017}.
\newblock \bibinfo{title}{Population-attributable risk proportion of clinical
  risk factors for breast cancer}.
\newblock \bibinfo{journal}{JAMA oncology} \bibinfo{volume}{3},
  \bibinfo{pages}{1228--1236}.
\bibitem[{Ferrari et~al.(2004)Ferrari, Rangayyan, Desautels, Borges and
  Frere}]{ferrari2004automatic}
\bibinfo{author}{Ferrari, R.J.}, \bibinfo{author}{Rangayyan, R.M.},
  \bibinfo{author}{Desautels, J.L.}, \bibinfo{author}{Borges, R.},
  \bibinfo{author}{Frere, A.F.}, \bibinfo{year}{2004}.
\newblock \bibinfo{title}{Automatic identification of the pectoral muscle in
  mammograms}.
\newblock \bibinfo{journal}{IEEE transactions on medical imaging}
  \bibinfo{volume}{23}, \bibinfo{pages}{232--245}.
\bibitem[{Freer(2015)}]{freer2015mammographic}
\bibinfo{author}{Freer, P.E.}, \bibinfo{year}{2015}.
\newblock \bibinfo{title}{Mammographic breast density: impact on breast cancer
  risk and implications for screening}.
\newblock \bibinfo{journal}{Radiographics} \bibinfo{volume}{35},
  \bibinfo{pages}{302--315}.
\bibitem[{Gastounioti et~al.(2018)Gastounioti, Hsieh, Cohen, Pantalone, Conant
  and Kontos}]{gastounioti2018incorporating}
\bibinfo{author}{Gastounioti, A.}, \bibinfo{author}{Hsieh, M.K.},
  \bibinfo{author}{Cohen, E.}, \bibinfo{author}{Pantalone, L.},
  \bibinfo{author}{Conant, E.F.}, \bibinfo{author}{Kontos, D.},
  \bibinfo{year}{2018}.
\newblock \bibinfo{title}{Incorporating breast anatomy in computational
  phenotyping of mammographic parenchymal patterns for breast cancer risk
  estimation}.
\newblock \bibinfo{journal}{Scientific reports} \bibinfo{volume}{8},
  \bibinfo{pages}{17489}.
\bibitem[{Gastounioti et~al.(2020)Gastounioti, Kasi, Scott, Brandt, Jensen,
  Hruska, Wu, Norman, Conant, Winham et~al.}]{gastounioti2020evaluation}
\bibinfo{author}{Gastounioti, A.}, \bibinfo{author}{Kasi, C.D.},
  \bibinfo{author}{Scott, C.G.}, \bibinfo{author}{Brandt, K.R.},
  \bibinfo{author}{Jensen, M.R.}, \bibinfo{author}{Hruska, C.B.},
  \bibinfo{author}{Wu, F.F.}, \bibinfo{author}{Norman, A.D.},
  \bibinfo{author}{Conant, E.F.}, \bibinfo{author}{Winham, S.J.}, et~al.,
  \bibinfo{year}{2020}.
\newblock \bibinfo{title}{Evaluation of libra software for fully automated
  mammographic density assessment in breast cancer risk prediction}.
\newblock \bibinfo{journal}{Radiology} , \bibinfo{pages}{192509}.
\bibitem[{Hamidinekoo et~al.(2018)Hamidinekoo, Denton, Rampun, Honnor and
  Zwiggelaar}]{hamidinekoo2018deep}
\bibinfo{author}{Hamidinekoo, A.}, \bibinfo{author}{Denton, E.},
  \bibinfo{author}{Rampun, A.}, \bibinfo{author}{Honnor, K.},
  \bibinfo{author}{Zwiggelaar, R.}, \bibinfo{year}{2018}.
\newblock \bibinfo{title}{Deep learning in mammography and breast histology, an
  overview and future trends}.
\newblock \bibinfo{journal}{Medical image analysis} \bibinfo{volume}{47},
  \bibinfo{pages}{45--67}.
\bibitem[{Hartman et~al.(2008)Hartman, Highnam, Warren and
  Jackson}]{hartman2008volumetric}
\bibinfo{author}{Hartman, K.}, \bibinfo{author}{Highnam, R.},
  \bibinfo{author}{Warren, R.}, \bibinfo{author}{Jackson, V.},
  \bibinfo{year}{2008}.
\newblock \bibinfo{title}{Volumetric assessment of breast tissue composition
  from ffdm images}, in: \bibinfo{booktitle}{International Workshop on Digital
  Mammography}, \bibinfo{organization}{Springer}. pp. \bibinfo{pages}{33--39}.
\bibitem[{Irshad et~al.(2016)Irshad, Leddy, Ackerman, Cluver, Pavic, Abid and
  Lewis}]{irshad2016effects}
\bibinfo{author}{Irshad, A.}, \bibinfo{author}{Leddy, R.},
  \bibinfo{author}{Ackerman, S.}, \bibinfo{author}{Cluver, A.},
  \bibinfo{author}{Pavic, D.}, \bibinfo{author}{Abid, A.},
  \bibinfo{author}{Lewis, M.C.}, \bibinfo{year}{2016}.
\newblock \bibinfo{title}{Effects of changes in bi-rads density assessment
  guidelines (fourth versus fifth edition) on breast density assessment:
  intra-and interreader agreements and density distribution}.
\newblock \bibinfo{journal}{American Journal of Roentgenology}
  \bibinfo{volume}{207}, \bibinfo{pages}{1366--1371}.
\bibitem[{Kallenberg et~al.(2016)Kallenberg, Petersen, Nielsen, Ng, Diao, Igel,
  Vachon, Holland, Winkel, Karssemeijer et~al.}]{kallenberg2016unsupervised}
\bibinfo{author}{Kallenberg, M.}, \bibinfo{author}{Petersen, K.},
  \bibinfo{author}{Nielsen, M.}, \bibinfo{author}{Ng, A.Y.},
  \bibinfo{author}{Diao, P.}, \bibinfo{author}{Igel, C.},
  \bibinfo{author}{Vachon, C.M.}, \bibinfo{author}{Holland, K.},
  \bibinfo{author}{Winkel, R.R.}, \bibinfo{author}{Karssemeijer, N.}, et~al.,
  \bibinfo{year}{2016}.
\newblock \bibinfo{title}{Unsupervised deep learning applied to breast density
  segmentation and mammographic risk scoring}.
\newblock \bibinfo{journal}{IEEE transactions on medical imaging}
  \bibinfo{volume}{35}, \bibinfo{pages}{1322--1331}.
\bibitem[{Kaul et~al.(2019)Kaul, Manandhar and Pears}]{kaul2019focusnet}
\bibinfo{author}{Kaul, C.}, \bibinfo{author}{Manandhar, S.},
  \bibinfo{author}{Pears, N.}, \bibinfo{year}{2019}.
\newblock \bibinfo{title}{Focusnet: An attention-based fully convolutional
  network for medical image segmentation}.
\newblock \bibinfo{journal}{arXiv preprint arXiv:1902.03091} .
\bibitem[{Keller et~al.(2012)Keller, Nathan, Wang, Zheng, Gee, Conant and
  Kontos}]{keller2012estimation}
\bibinfo{author}{Keller, B.M.}, \bibinfo{author}{Nathan, D.L.},
  \bibinfo{author}{Wang, Y.}, \bibinfo{author}{Zheng, Y.},
  \bibinfo{author}{Gee, J.C.}, \bibinfo{author}{Conant, E.F.},
  \bibinfo{author}{Kontos, D.}, \bibinfo{year}{2012}.
\newblock \bibinfo{title}{Estimation of breast percent density in raw and
  processed full field digital mammography images via adaptive fuzzy c-means
  clustering and support vector machine segmentation}.
\newblock \bibinfo{journal}{Medical physics} \bibinfo{volume}{39},
  \bibinfo{pages}{4903--4917}.
\bibitem[{Kontos and Conant(2019)}]{kontos2019can}
\bibinfo{author}{Kontos, D.}, \bibinfo{author}{Conant, E.F.},
  \bibinfo{year}{2019}.
\newblock \bibinfo{title}{Can ai help make screening mammography “lean”?}
\bibitem[{Kooi et~al.(2017)Kooi, Litjens, Van~Ginneken, Gubern-M{\'e}rida,
  S{\'a}nchez, Mann, den Heeten and Karssemeijer}]{kooi2017large}
\bibinfo{author}{Kooi, T.}, \bibinfo{author}{Litjens, G.},
  \bibinfo{author}{Van~Ginneken, B.}, \bibinfo{author}{Gubern-M{\'e}rida, A.},
  \bibinfo{author}{S{\'a}nchez, C.I.}, \bibinfo{author}{Mann, R.},
  \bibinfo{author}{den Heeten, A.}, \bibinfo{author}{Karssemeijer, N.},
  \bibinfo{year}{2017}.
\newblock \bibinfo{title}{Large scale deep learning for computer aided
  detection of mammographic lesions}.
\newblock \bibinfo{journal}{Medical image analysis} \bibinfo{volume}{35},
  \bibinfo{pages}{303--312}.
\bibitem[{Kwok et~al.(2004)Kwok, Chandrasekhar, Attikiouzel and
  Rickard}]{kwok2004automatic}
\bibinfo{author}{Kwok, S.M.}, \bibinfo{author}{Chandrasekhar, R.},
  \bibinfo{author}{Attikiouzel, Y.}, \bibinfo{author}{Rickard, M.T.},
  \bibinfo{year}{2004}.
\newblock \bibinfo{title}{Automatic pectoral muscle segmentation on
  mediolateral oblique view mammograms}.
\newblock \bibinfo{journal}{IEEE transactions on medical imaging}
  \bibinfo{volume}{23}, \bibinfo{pages}{1129--1140}.
\bibitem[{Lehman et~al.(2018)Lehman, Yala, Schuster, Dontchos, Bahl, Swanson
  and Barzilay}]{lehman2018mammographic}
\bibinfo{author}{Lehman, C.D.}, \bibinfo{author}{Yala, A.},
  \bibinfo{author}{Schuster, T.}, \bibinfo{author}{Dontchos, B.},
  \bibinfo{author}{Bahl, M.}, \bibinfo{author}{Swanson, K.},
  \bibinfo{author}{Barzilay, R.}, \bibinfo{year}{2018}.
\newblock \bibinfo{title}{Mammographic breast density assessment using deep
  learning: clinical implementation}.
\newblock \bibinfo{journal}{Radiology} \bibinfo{volume}{290},
  \bibinfo{pages}{52--58}.
\bibitem[{Li et~al.(2013)Li, Chen, Yang and Yang}]{li2013pectoral}
\bibinfo{author}{Li, Y.}, \bibinfo{author}{Chen, H.}, \bibinfo{author}{Yang,
  Y.}, \bibinfo{author}{Yang, N.}, \bibinfo{year}{2013}.
\newblock \bibinfo{title}{Pectoral muscle segmentation in mammograms based on
  homogenous texture and intensity deviation}.
\newblock \bibinfo{journal}{Pattern Recognition} \bibinfo{volume}{46},
  \bibinfo{pages}{681--691}.
\bibitem[{Ma et~al.(2019)Ma, Wei, Zhou, Helvie, Chan, Hadjiiski and
  Lu}]{ma2019automated}
\bibinfo{author}{Ma, X.}, \bibinfo{author}{Wei, J.}, \bibinfo{author}{Zhou,
  C.}, \bibinfo{author}{Helvie, M.A.}, \bibinfo{author}{Chan, H.P.},
  \bibinfo{author}{Hadjiiski, L.M.}, \bibinfo{author}{Lu, Y.},
  \bibinfo{year}{2019}.
\newblock \bibinfo{title}{Automated pectoral muscle identification on mlo-view
  mammograms: Comparison of deep neural network to conventional computer
  vision}.
\newblock \bibinfo{journal}{Medical physics} \bibinfo{volume}{46},
  \bibinfo{pages}{2103--2114}.
\bibitem[{Maghsoudi et~al.(2020)Maghsoudi, Gastounioti, Pantalone, Davatzikos,
  Bakas and Kontos}]{maghsoudi2020net}
\bibinfo{author}{Maghsoudi, O.H.}, \bibinfo{author}{Gastounioti, A.},
  \bibinfo{author}{Pantalone, L.}, \bibinfo{author}{Davatzikos, C.},
  \bibinfo{author}{Bakas, S.}, \bibinfo{author}{Kontos, D.},
  \bibinfo{year}{2020}.
\newblock \bibinfo{title}{O-net: An overall convolutional network for
  segmentation tasks}, in: \bibinfo{booktitle}{International Workshop on
  Machine Learning in Medical Imaging}, \bibinfo{organization}{Springer}. pp.
  \bibinfo{pages}{199--209}.
\bibitem[{McCarthy et~al.(2016)McCarthy, Keller, Pantalone, Hsieh, Synnestvedt,
  Conant, Armstrong and Kontos}]{mccarthy2016racial}
\bibinfo{author}{McCarthy, A.M.}, \bibinfo{author}{Keller, B.M.},
  \bibinfo{author}{Pantalone, L.M.}, \bibinfo{author}{Hsieh, M.K.},
  \bibinfo{author}{Synnestvedt, M.}, \bibinfo{author}{Conant, E.F.},
  \bibinfo{author}{Armstrong, K.}, \bibinfo{author}{Kontos, D.},
  \bibinfo{year}{2016}.
\newblock \bibinfo{title}{Racial differences in quantitative measures of area
  and volumetric breast density}.
\newblock \bibinfo{journal}{Journal of the National Cancer Institute}
  \bibinfo{volume}{108}, \bibinfo{pages}{djw104}.
\bibitem[{Mohamed et~al.(2018)Mohamed, Berg, Peng, Luo, Jankowitz and
  Wu}]{mohamed2018deep}
\bibinfo{author}{Mohamed, A.A.}, \bibinfo{author}{Berg, W.A.},
  \bibinfo{author}{Peng, H.}, \bibinfo{author}{Luo, Y.},
  \bibinfo{author}{Jankowitz, R.C.}, \bibinfo{author}{Wu, S.},
  \bibinfo{year}{2018}.
\newblock \bibinfo{title}{A deep learning method for classifying mammographic
  breast density categories}.
\newblock \bibinfo{journal}{Medical physics} \bibinfo{volume}{45},
  \bibinfo{pages}{314--321}.
\bibitem[{Mortazi and Bagci(2018)}]{mortazi2018automatically}
\bibinfo{author}{Mortazi, A.}, \bibinfo{author}{Bagci, U.},
  \bibinfo{year}{2018}.
\newblock \bibinfo{title}{Automatically designing cnn architectures for medical
  image segmentation}, in: \bibinfo{booktitle}{International Workshop on
  Machine Learning in Medical Imaging}, \bibinfo{organization}{Springer}. pp.
  \bibinfo{pages}{98--106}.
\bibitem[{Murugesan et~al.(2019)Murugesan, Sarveswaran, Shankaranarayana, Ram
  and Sivaprakasam}]{murugesan2019psi}
\bibinfo{author}{Murugesan, B.}, \bibinfo{author}{Sarveswaran, K.},
  \bibinfo{author}{Shankaranarayana, S.M.}, \bibinfo{author}{Ram, K.},
  \bibinfo{author}{Sivaprakasam, M.}, \bibinfo{year}{2019}.
\newblock \bibinfo{title}{Psi-net: Shape and boundary aware joint multi-task
  deep network for medical image segmentation}.
\newblock \bibinfo{journal}{arXiv preprint arXiv:1902.04099} .
\bibitem[{Mustra and Grgic(2013)}]{mustra2013robust}
\bibinfo{author}{Mustra, M.}, \bibinfo{author}{Grgic, M.},
  \bibinfo{year}{2013}.
\newblock \bibinfo{title}{Robust automatic breast and pectoral muscle
  segmentation from scanned mammograms}.
\newblock \bibinfo{journal}{Signal processing} \bibinfo{volume}{93},
  \bibinfo{pages}{2817--2827}.
\bibitem[{Mustra et~al.(2016)Mustra, Grgic and Rangayyan}]{mustra2016review}
\bibinfo{author}{Mustra, M.}, \bibinfo{author}{Grgic, M.},
  \bibinfo{author}{Rangayyan, R.M.}, \bibinfo{year}{2016}.
\newblock \bibinfo{title}{Review of recent advances in segmentation of the
  breast boundary and the pectoral muscle in mammograms}.
\newblock \bibinfo{journal}{Medical \& biological engineering \& computing}
  \bibinfo{volume}{54}, \bibinfo{pages}{1003--1024}.
\bibitem[{Nagi et~al.(2010)Nagi, Kareem, Nagi and Ahmed}]{nagi2010automated}
\bibinfo{author}{Nagi, J.}, \bibinfo{author}{Kareem, S.A.},
  \bibinfo{author}{Nagi, F.}, \bibinfo{author}{Ahmed, S.K.},
  \bibinfo{year}{2010}.
\newblock \bibinfo{title}{Automated breast profile segmentation for roi
  detection using digital mammograms}, in: \bibinfo{booktitle}{2010 IEEE EMBS
  conference on biomedical engineering and sciences (IECBES)},
  \bibinfo{organization}{IEEE}. pp. \bibinfo{pages}{87--92}.
\bibitem[{Rampun et~al.(2017)Rampun, Morrow, Scotney and
  Winder}]{rampun2017fully}
\bibinfo{author}{Rampun, A.}, \bibinfo{author}{Morrow, P.J.},
  \bibinfo{author}{Scotney, B.W.}, \bibinfo{author}{Winder, J.},
  \bibinfo{year}{2017}.
\newblock \bibinfo{title}{Fully automated breast boundary and pectoral muscle
  segmentation in mammograms}.
\newblock \bibinfo{journal}{Artificial intelligence in medicine}
  \bibinfo{volume}{79}, \bibinfo{pages}{28--41}.
\bibitem[{Regini et~al.(2014)Regini, Mariscotti, Durando, Ghione, Luparia,
  Campanino, Bianchi, Bergamasco, Fonio and Gandini}]{regini2014radiological}
\bibinfo{author}{Regini, E.}, \bibinfo{author}{Mariscotti, G.},
  \bibinfo{author}{Durando, M.}, \bibinfo{author}{Ghione, G.},
  \bibinfo{author}{Luparia, A.}, \bibinfo{author}{Campanino, P.P.},
  \bibinfo{author}{Bianchi, C.C.}, \bibinfo{author}{Bergamasco, L.},
  \bibinfo{author}{Fonio, P.}, \bibinfo{author}{Gandini, G.},
  \bibinfo{year}{2014}.
\newblock \bibinfo{title}{Radiological assessment of breast density by visual
  classification (bi--rads) compared to automated volumetric digital software
  (quantra): implications for clinical practice}.
\newblock \bibinfo{journal}{La radiologia medica} \bibinfo{volume}{119},
  \bibinfo{pages}{741--749}.
\bibitem[{Rodr{\'\i}guez-Ruiz et~al.(2018)Rodr{\'\i}guez-Ruiz, Krupinski,
  Mordang, Schilling, Heywang-K{\"o}brunner, Sechopoulos and
  Mann}]{rodriguez2018detection}
\bibinfo{author}{Rodr{\'\i}guez-Ruiz, A.}, \bibinfo{author}{Krupinski, E.},
  \bibinfo{author}{Mordang, J.J.}, \bibinfo{author}{Schilling, K.},
  \bibinfo{author}{Heywang-K{\"o}brunner, S.H.}, \bibinfo{author}{Sechopoulos,
  I.}, \bibinfo{author}{Mann, R.M.}, \bibinfo{year}{2018}.
\newblock \bibinfo{title}{Detection of breast cancer with mammography: effect
  of an artificial intelligence support system}.
\newblock \bibinfo{journal}{Radiology} \bibinfo{volume}{290},
  \bibinfo{pages}{305--314}.
\bibitem[{Rodriguez-Ruiz et~al.(2018)Rodriguez-Ruiz, Teuwen, Chung,
  Karssemeijer, Chevalier, Gubern-Merida and
  Sechopoulos}]{rodriguez2018pectoral}
\bibinfo{author}{Rodriguez-Ruiz, A.}, \bibinfo{author}{Teuwen, J.},
  \bibinfo{author}{Chung, K.}, \bibinfo{author}{Karssemeijer, N.},
  \bibinfo{author}{Chevalier, M.}, \bibinfo{author}{Gubern-Merida, A.},
  \bibinfo{author}{Sechopoulos, I.}, \bibinfo{year}{2018}.
\newblock \bibinfo{title}{Pectoral muscle segmentation in breast tomosynthesis
  with deep learning}, in: \bibinfo{booktitle}{Medical Imaging 2018:
  Computer-Aided Diagnosis}, \bibinfo{organization}{International Society for
  Optics and Photonics}. p. \bibinfo{pages}{105752J}.
\bibitem[{Ronneberger et~al.(2015)Ronneberger, Fischer and
  Brox}]{ronneberger2015u}
\bibinfo{author}{Ronneberger, O.}, \bibinfo{author}{Fischer, P.},
  \bibinfo{author}{Brox, T.}, \bibinfo{year}{2015}.
\newblock \bibinfo{title}{U-net: Convolutional networks for biomedical image
  segmentation}, in: \bibinfo{booktitle}{International Conference on Medical
  image computing and computer-assisted intervention},
  \bibinfo{organization}{Springer}. pp. \bibinfo{pages}{234--241}.
\bibitem[{Rueden et~al.(2017)Rueden, Schindelin, Hiner, DeZonia, Walter, Arena
  and Eliceiri}]{rueden2017imagej2}
\bibinfo{author}{Rueden, C.T.}, \bibinfo{author}{Schindelin, J.},
  \bibinfo{author}{Hiner, M.C.}, \bibinfo{author}{DeZonia, B.E.},
  \bibinfo{author}{Walter, A.E.}, \bibinfo{author}{Arena, E.T.},
  \bibinfo{author}{Eliceiri, K.W.}, \bibinfo{year}{2017}.
\newblock \bibinfo{title}{Imagej2: Imagej for the next generation of scientific
  image data}.
\newblock \bibinfo{journal}{BMC bioinformatics} \bibinfo{volume}{18},
  \bibinfo{pages}{529}.
\bibitem[{Shi et~al.(2018)Shi, Zhong, Rampun and Wang}]{shi2018hierarchical}
\bibinfo{author}{Shi, P.}, \bibinfo{author}{Zhong, J.},
  \bibinfo{author}{Rampun, A.}, \bibinfo{author}{Wang, H.},
  \bibinfo{year}{2018}.
\newblock \bibinfo{title}{A hierarchical pipeline for breast boundary
  segmentation and calcification detection in mammograms}.
\newblock \bibinfo{journal}{Computers in biology and medicine}
  \bibinfo{volume}{96}, \bibinfo{pages}{178--188}.
\bibitem[{Sprague et~al.(2016)Sprague, Conant, Onega, Garcia, Beaber,
  Herschorn, Lehman, Tosteson, Lacson, Schnall et~al.}]{sprague2016variation}
\bibinfo{author}{Sprague, B.L.}, \bibinfo{author}{Conant, E.F.},
  \bibinfo{author}{Onega, T.}, \bibinfo{author}{Garcia, M.P.},
  \bibinfo{author}{Beaber, E.F.}, \bibinfo{author}{Herschorn, S.D.},
  \bibinfo{author}{Lehman, C.D.}, \bibinfo{author}{Tosteson, A.N.},
  \bibinfo{author}{Lacson, R.}, \bibinfo{author}{Schnall, M.D.}, et~al.,
  \bibinfo{year}{2016}.
\newblock \bibinfo{title}{Variation in mammographic breast density assessments
  among radiologists in clinical practice: a multicenter observational study}.
\newblock \bibinfo{journal}{Annals of internal medicine} \bibinfo{volume}{165},
  \bibinfo{pages}{457--464}.
\bibitem[{Szegedy et~al.(2017)Szegedy, Ioffe, Vanhoucke and
  Alemi}]{szegedy2017inception}
\bibinfo{author}{Szegedy, C.}, \bibinfo{author}{Ioffe, S.},
  \bibinfo{author}{Vanhoucke, V.}, \bibinfo{author}{Alemi, A.A.},
  \bibinfo{year}{2017}.
\newblock \bibinfo{title}{Inception-v4, inception-resnet and the impact of
  residual connections on learning}, in: \bibinfo{booktitle}{Thirty-First AAAI
  Conference on Artificial Intelligence}.
\bibitem[{Taghanaki et~al.(2017)Taghanaki, Liu, Miles and
  Hamarneh}]{taghanaki2017geometry}
\bibinfo{author}{Taghanaki, S.A.}, \bibinfo{author}{Liu, Y.},
  \bibinfo{author}{Miles, B.}, \bibinfo{author}{Hamarneh, G.},
  \bibinfo{year}{2017}.
\newblock \bibinfo{title}{Geometry-based pectoral muscle segmentation from mlo
  mammogram views}.
\newblock \bibinfo{journal}{IEEE Transactions on Biomedical Engineering}
  \bibinfo{volume}{64}, \bibinfo{pages}{2662--2671}.
\bibitem[{Van~Griethuysen et~al.(2017)Van~Griethuysen, Fedorov, Parmar, Hosny,
  Aucoin, Narayan, Beets-Tan, Fillion-Robin, Pieper and
  Aerts}]{van2017computational}
\bibinfo{author}{Van~Griethuysen, J.J.}, \bibinfo{author}{Fedorov, A.},
  \bibinfo{author}{Parmar, C.}, \bibinfo{author}{Hosny, A.},
  \bibinfo{author}{Aucoin, N.}, \bibinfo{author}{Narayan, V.},
  \bibinfo{author}{Beets-Tan, R.G.}, \bibinfo{author}{Fillion-Robin, J.C.},
  \bibinfo{author}{Pieper, S.}, \bibinfo{author}{Aerts, H.J.},
  \bibinfo{year}{2017}.
\newblock \bibinfo{title}{Computational radiomics system to decode the
  radiographic phenotype}.
\newblock \bibinfo{journal}{Cancer research} \bibinfo{volume}{77},
  \bibinfo{pages}{e104--e107}.
\bibitem[{Wang et~al.(2016)Wang, Yang, Cai, Tan, Jin and
  Li}]{wang2016discrimination}
\bibinfo{author}{Wang, J.}, \bibinfo{author}{Yang, X.}, \bibinfo{author}{Cai,
  H.}, \bibinfo{author}{Tan, W.}, \bibinfo{author}{Jin, C.},
  \bibinfo{author}{Li, L.}, \bibinfo{year}{2016}.
\newblock \bibinfo{title}{Discrimination of breast cancer with
  microcalcifications on mammography by deep learning}.
\newblock \bibinfo{journal}{Scientific reports} \bibinfo{volume}{6},
  \bibinfo{pages}{27327}.
\bibitem[{Yala et~al.(2019)Yala, Schuster, Miles, Barzilay and
  Lehman}]{yala2019deep}
\bibinfo{author}{Yala, A.}, \bibinfo{author}{Schuster, T.},
  \bibinfo{author}{Miles, R.}, \bibinfo{author}{Barzilay, R.},
  \bibinfo{author}{Lehman, C.}, \bibinfo{year}{2019}.
\newblock \bibinfo{title}{A deep learning model to triage screening mammograms:
  a simulation study}.
\newblock \bibinfo{journal}{Radiology} , \bibinfo{pages}{182908}.
\bibitem[{Zhou et~al.(2001)Zhou, Chan, Petrick, Helvie, Goodsitt, Sahiner and
  Hadjiiski}]{zhou2001computerized}
\bibinfo{author}{Zhou, C.}, \bibinfo{author}{Chan, H.P.},
  \bibinfo{author}{Petrick, N.}, \bibinfo{author}{Helvie, M.A.},
  \bibinfo{author}{Goodsitt, M.M.}, \bibinfo{author}{Sahiner, B.},
  \bibinfo{author}{Hadjiiski, L.M.}, \bibinfo{year}{2001}.
\newblock \bibinfo{title}{Computerized image analysis: estimation of breast
  density on mammograms}.
\newblock \bibinfo{journal}{Medical physics} \bibinfo{volume}{28},
  \bibinfo{pages}{1056--1069}.

\end{thebibliography}

\section*{Supplementary Material}
Please find Supplementary Material on \href{https://upenn.box.com/s/9myv3at6cu3kzqsuphv56uwstj0fhoh9}{GitHub} (it will be moved to Github).

\end{document}